\documentclass[12pt]{article}
\pdfoutput=1
\usepackage{epsf,amsfonts,amssymb,epsfig,amsmath,mathtools,graphics,slashed}
\usepackage{hyperref,hep,graphicx,subfig}
\usepackage{rotating}

\newcommand{\nn}{\notag \\}

\makeatletter

\@addtoreset{equation}{section}
\makeatother

\begin{document}

\begin{titlepage}

\vfill

\begin{flushright}
Imperial/TP/2018/JG/01\\
DCPT-17/29
\end{flushright}

\vfill

\begin{center}
   \baselineskip=16pt
   {\Large\bf Incoherent transport for phases that spontaneously break translations}
  \vskip 1.5cm
  \vskip 1.5cm
Aristomenis Donos$^1$, Jerome P. Gauntlett$^2$, Tom Griffin$^2$ and Vaios Ziogas$^1$\\
     \vskip .6cm
     \begin{small}
      \textit{$^1$Centre for Particle Theory and Department of Mathematical Sciences\\Durham University, Durham, DH1 3LE, U.K.}
        \end{small}\\    
         \begin{small}\vskip .6cm
      \textit{$^2$Blackett Laboratory, 
  Imperial College\\ London, SW7 2AZ, U.K.}
        \end{small}\\
        \end{center}
     \vskip .6cm
\vfill

\begin{center}
\textbf{Abstract}
\end{center}
\begin{quote}
We consider phases of matter at finite charge density which spontaneously break spatial translations. Without taking
a hydrodynamic limit we identify a boost invariant incoherent current operator. We also derive expressions for the small frequency 
behaviour of the thermoelectric conductivities generalising those that have been derived in a translationally invariant context.
Within holographic constructions we show that the DC conductivity for the incoherent current can be obtained from a solution to a Stokes flow for an auxiliary fluid on the black hole horizon combined with specific thermodynamic quantities associated with the equilibrium black hole solutions. 
\end{quote}

\vfill

\end{titlepage}

\setcounter{equation}{0}
\section{Introduction}
Studying the thermoelectric transport properties of quantum critical states of matter at finite charge density is 
a topic of great theoretical and practical importance. For `clean systems', i.e. systems that are translationally invariant
and hence without a mechanism for momentum to dissipate, it is well known that the DC conductivities are infinite. More
precisely, the translation invariance implies that momentum is conserved and this leads to the appearance of
a delta function in the thermoelectric AC conductivities at zero frequency.

For translationally invariant systems the notion of an `incoherent current' was introduced in \cite{Davison:2015taa},
building on \cite{Davison:2015bea}.
This was defined to be a linear combination of the electric and heat currents that has zero overlap with the momentum operator. 
Using the hydrodynamic results of \cite{Hartnoll:2007ih}, which implicitly assumed that the system was not in a superfluid state, it is then
easy to see that the incoherent current should have finite DC conductivity, $[\sigma_{inc}]_{DC}$. 
In first order relativistic hydrodynamics, there is only one independent transport coefficient for both neutral systems as well as systems at finite chemical potential. In this context, the retarded two point function of the incoherent current operator provides a generalisation of the Kubo formula for that transport coefficient which is
appropriate for systems at finite chemical potential. 
It was shown in \cite{Davison:2015taa} that
expressions for the low frequency behaviour of the thermoelectric conductivities can be expressed in terms of $[\sigma_{inc}]_{DC}$
and certain thermodynamics quantities. Furthermore, it was also shown how $[\sigma_{inc}]_{DC}$
can be calculated within a specific class of holographic models from data at the black hole horizon.

The goal of this short paper is to generalise some of these results to phases of 
relativistic systems, held at finite chemical potential with respect to an abelian
symmetry, that break translations spontaneously. Our general arguments, which are rather simple, will not assume any hydrodynamic limit of the system. That is, for a given temperature we will allow for phases with arbitrary
spatial modulation. We will identify a universal boost-invariant incoherent current and argue that when there is no superfluid 
the low frequency  behaviour of the thermoelectric conductivities can still be expressed in terms of
certain thermodynamics quantities as well as the finite incoherent DC conductivity, $[\sigma_{inc}]_{DC}$. Within a holographic context, describing a strongly coupled system, we also explain how 
$[\sigma_{inc}]_{DC}$ can be calculated in terms of a Stokes flow on the spatially modulated  black hole horizon, supplemented with some thermodynamic quantities of the background. This extends the results of
\cite{Donos:2015gia} that obtained the DC conductivities for holographic systems for which the translations are explicitly broken.

Naturally, we will focus on the properties of spatially modulated phases that are thermodynamically preferred. Such phases, which may have anisotropic spatial modulation,
necessarily satisfy the condition $\langle \bar{T}^{ij} \rangle =p\delta^{ij}$, where $\bar{T}^{ij}$ is the constant zero mode part of the
spatial components of
the stress tensor \cite{Donos:2013cka,Donos:2015eew}. However, since spatially modulated phases 
in which this condition is not satisfied have been analysed in a holographic context in 
\cite{Amoretti:2017frz} we briefly comment on some of the modified formula 
in appendix \ref{nonthermpref}. In particular our general results on how to derive the $[\sigma_{inc}]_{DC}$ within holography immediately lead to the result
presented in \cite{Amoretti:2017frz} for the specific holographic model studied there.

More generally, charge and spin density waves and their impact on the phenomenology of condensed matter systems have been of central interest for a long time e.g. \cite{Gruner:1988zz}. Some more recent work on thermoelectric transport for phases that spontaneously break translations has appeared in
\cite{Delacretaz:2017zxd}, which included the effects of disorder and pinning in a hydrodynamic description, as well as in a number of
holographic studies, including \cite{Andrade:2017cnc,Alberte:2017oqx,Amoretti:2017axe} 
and brane probe models \cite{Jokela:2016xuy, Jokela:2017ltu}. 
An interesting open topic, which is left for the future, would be to derive the effective hydrodynamic description of 
the specific examples of spontaneously formed density wave states which have already been studied within holography, 
along the lines of \cite{Blake:2015epa}.

\section{Boost invariant incoherent current}
Consider a relativistic quantum field theory at finite temperature defined on flat spacetime. We will consider the system to
be held at constant chemical potential, $\mu$, with respect to an abelian global symmetry. We will also allow
for the possibility for additional deformations of the Hamiltonian by a scalar operator $\mathcal{O}_\phi$ that
is parametrised by the constant source $\phi_{s}$. 
If $\mathcal{O}_\phi$ is odd under time reversal invariance then 
a non-zero $\phi_{s}$ will explicitly break time reversal invariance.

We are particularly interested in phases in which spatial translations
are broken spontaneously, but our analysis will also cover translationally invariant phases.
We will assume that the system reaches local thermodynamic equilibrium satisfying periodic boundary conditions 
generated by a set of lattice vectors $\{{\bf L}_i\}$. Thus, the expectation values of the
stress tensor density, $\langle T^{\mu\nu}\rangle$, the conserved abelian current density, $\langle J^\mu\rangle$, as well $\langle \mathcal{O}_\phi\rangle$ are all functions of the spatial coordinates, ${\bf x}$, which are taken to be cartesian coordinates, that
are invariant under shifts by any of the lattice vectors.
For any such function, $A{(\bf x})$, the zero mode is denoted by $\bar A$, with $\bar A=\oint A\equiv \frac{1}{vol}\int_{\{0\}}^{\{{\bf L_i}\} }d{\bf x}A{(\bf x})$, where the volume of a unit cell of the lattice is $vol\equiv \int_{\{0\}}^{\{{\bf L_i}\} }d{\bf x}$.

It is important to recall that the thermodynamically preferred configurations will necessarily satisfy certain constraints on the zero modes of these expectation values \cite{Donos:2013cka,Donos:2015eew}. In particular, by ensuring that the free energy is minimised over the moduli space of spontaneously generated lattices, we must have
$\langle \bar{T}^{ij} \rangle\equiv t^{ij} =p\delta^{ij}$, where $p$ is the spatially averaged constant pressure density and is related to the free energy density,
$w$, via $w=-p$. Defining the total charge density $\rho\equiv \langle \bar J^t \rangle$, the total energy density $\varepsilon\equiv -\langle \bar T^{t}{}_{t}\rangle$, and the total entropy density $s$, we also have the fundamental thermodynamic relation
$Ts+\rho\mu=\varepsilon+p$. It was also shown in \cite{Donos:2013cka} that the zero mode of the heat current must vanish, $\langle \bar Q^{i}\rangle=0$, where we recall that $Q^i\equiv -T^i{}_t- \mu J^{i}$. 
If the global $U(1)$ symmetry is not spontaneously broken, which will be the principle focus of this paper, 
by extending the arguments of \cite{Donos:2013cka}, we can invoke invariance under large gauge transformations with gauge parameter $\Lambda=x^{i}\,q_{i}$
to argue that $\langle \bar{J}^{i}\rangle=0$ as well. On the other hand for a superfluid one can have $\langle \bar Q^{i}\rangle=0$ with 
$\langle \bar J^{i}\rangle\ne  0$ since a non-trivial external gauge field of the form $A_{i}=q_{i}$ cannot be gauged away, being associated
with a supercurrent. However, for the thermodynamically preferred phase obtained by minimising the free energy with respect to  $q_{i}$, we have once again that $\langle \bar J^{i}\rangle=  0$. We also note here that $P_{(i)}\equiv \bar T^t{}_i$ is the time independent charge associated with the total momentum density operator in the $i$th direction.

We now deduce some simple facts about the two-point functions for the current-current retarded Green's functions. These can be obtained from Ward identities, generalising \cite{Herzog:2009xv}, but
we find it illuminating to obtain them by generating
a time-dependent perturbation via the coordinate transformation 
\begin{align}\label{boost}
x^{i}\to x^{i}+\lambda e^{-i\omega t}\,\xi^{i}\,,
\end{align}
where $\lambda$ is a small parameter and $\xi^i$ is a constant vector. 
Notice that for small $\omega$ this is a translation combined with a Galilean boost.
By taking the Lie derivative with respect to the vector $k^\mu=(0,\lambda e^{-i\omega t}\,\xi^{i})$, we
easily determine how various quantities transform. The transformed metric is $ds^2=-dt^2+\delta_{ij}dx^i dx^j
-2i\omega\,\lambda e^{-i\omega t}\,\xi_{i} dx^i dt$ and the 
perturbation $\delta g_{ti}=-i\omega\,\lambda e^{-i\omega t}\,\xi_{i}$ parametrises a source for the operator $T^{ti}$ in the action.
Equivalently, it generates\footnote{It also can be viewed as generating a source in the Hamiltonian for the operator 
$T^t{}_i$ with parameter $+{i\omega}\lambda e^{-i\omega t}\xi^{i}$.} 
a spatially independent source in the Hamiltonian associated with the
operator $-T^i{}_t=Q^i+\mu J^i$ and with parameter $+{i\omega}\lambda e^{-i\omega t}\xi_{i}$.

The coordinate transformation also modifies the stress tensor and current densities and we find 
\begin{align}\label{blippy}
\delta  T^t{}_{t} =&\lambda e^{-i\omega t}\left(\xi^k\partial_k T^t{}_{t}+i\omega\xi_i T^i{}_{t}\right)\,,
\qquad
\delta  T^t{}_{i} =\lambda e^{-i\omega t}\xi^k\partial_k T^t{}_{i}\,,
\nn
\delta  T^{i}{}_{t}=&\lambda e^{-i\omega t}\left(\xi^k\partial_k T^i{}_{t}+i\omega\,\left[\xi^{i}\, T^{t}{}_{t}- \xi^{j} T^{i}{}_j\right]\right)\,,
\nn
\delta  T^i{}_{j} =&\lambda e^{-i\omega t}\left(\xi^k\partial_k T^i{}_{j}+i\omega\xi^i T^t{}_{j}\right)\,,\qquad
\delta  \mathcal{O}_{\phi}=\lambda e^{-i\omega t}\,\xi^k\partial_k \mathcal{O}_{\phi}\,,\nn
\delta  J^{t} =&\lambda e^{-i\omega t}\,\xi^k\partial_k J^{t}\,,\qquad
\delta  J^{i}=\lambda e^{-i\omega t}\left(\xi^k\partial_k J^i+
i\omega\xi^i J^{t}\right)\,.
\end{align}
Focussing now on the zero modes we have 
\begin{align}\label{blippys}
\delta  \bar T^t{}_{t} =&
\lambda e^{-i\omega t}i\omega\xi_i \bar T^i{}_{t}\,,\qquad
\delta  \bar T^t{}_{i} =0\,,\nn
\delta  \bar T^{i}{}_{t}=&\lambda e^{-i\omega t}i\omega\,\left[\xi^{i}\, \bar T^{t}{}_{t}- \xi^{j} \bar T^{i}{}_{j}\right]\,,\qquad
\delta \bar  T^i{}_{j} =\lambda e^{-i\omega t}i\omega\xi^i \bar T^t{}_{j}\,,\nn
\delta  \bar J^{t} =&0\,,\qquad
\delta  \bar J^{i}=\lambda e^{-i\omega t}i\omega \bar J^t\,\xi^{i}\,,
\qquad
\delta  \bar{\mathcal{O}}_{\phi}=0\,.
\end{align}
In particular, we notice from the first line that while the Hamiltonian has changed, the total momentum density operator is unchanged $\delta P_{(i)}\equiv \delta  \bar T^t{}_{i}=0$ (and it is worth highlighting that $\delta  \bar T^{i}{}_{t}\ne0$).

From these expressions we can immediately read off the one-point function responses of the system to the source for the operator
$Q^i+\mu J^i$, with parameter $+{i\omega}\lambda e^{-i\omega t}\xi_{i}$. For example, we have 
\begin{align}\label{blippys2}
\delta \langle \bar T^{i}{}_{t}\rangle=&-\lambda e^{-i\omega t}i\omega \xi^i(\varepsilon +p)\,,\qquad
\delta \langle \bar J^{i}\rangle=\lambda e^{-i\omega t}i\omega \rho\,\xi^{i}\,.
\end{align}
Hence, we can immediately deduce, in particular, that 
\begin{align}\label{eefive}
G_{J^{i}{(Q^j+\mu J^j)}}(\omega,{\bf 0})={\rho}\,\delta^{ij}\,,\qquad G_{{(Q^i+\mu J^i)}{(Q^j+\mu J^j)}}(\omega,{\bf 0})=
(\varepsilon +p)\delta^{ij}\,\,.
\end{align}
Here we are using the notation for the retarded Green's functions discussed in \cite{Donos:2017gej}, with $G_{AB}(\omega,{\bf k})$ determining
the zero mode linear response of an operator $A$ to the application of a source for the operator $B$ parametrised
by a single Fourier mode labelled by $(\omega,{\bf k})$. 

We can obtain further information using Onsager's relations, which relate Green's functions in 
a given background to those in a background with time-reversed sources. In the set-ups we are considering the only possible source 
that breaks time reversal invariance is the scalar source $\phi_s$ in the particular case 
when the operator $\mathcal{O}$ is odd under time-reversal. Thus, for example, we have in general $G_{J^iQ^j}(\omega,{\bf 0})
=G'_{Q^jJ^i}(\omega,{\bf 0})$, where the prime denotes the background with the opposite sign for $\phi_s$.
Now, suppose that we consider the time reversed background
and then carry out exactly the same transformations as above.
We then deduce the results \eqref{eefive} for the primed Green's functions, i.e. in the time reversed background, 
with exactly the same right hand sides (since they are inert under time-reversal):
$G'_{J^{i}{(Q^j+\mu J^j)}}(\omega,{\bf 0})={\rho}\,\delta^{ij}$ and  $G'_{{(Q^i+\mu J^i)}{(Q^j+\mu J^j)}}(\omega,{\bf 0})=
(\varepsilon +p)\delta^{ij}$.
Using Onsager's relations on these expressions, and that they are explicitly symmetric in $i$ and $j$,
we can deduce that for the Green's functions in the original background, in addition to \eqref{eefive} we also have
\begin{align}\label{eefive2}
G_{(Q^i+\mu J^i){J}^{j}}(\omega,{\bf 0})=&{\rho}\,\delta^{ij}\,.
\end{align}
As a corollary, we also have, in the original background, $G_{J^{i}{Q^j}}(\omega,{\bf 0})=G_{Q^{i}{J^j}}(\omega,{\bf 0})$.

After multiplying \eqref{eefive}, \eqref{eefive2} by $i/\omega$ we then deduce the following relations for the thermoelectric AC conductivity
\begin{align}\label{poles}
\mu\,\sigma^{ij}(\omega)+T{\alpha}^{ij}(\omega)=&\frac{i{\rho}}{\omega}\,\delta^{ij}\,,\nn
\mu T\bar\alpha^{ij}(\omega)+T\bar{\kappa}^{ij}(\omega)=&\frac{iTs}{\omega}\,\delta^{ij}\,,
\end{align}
and we also have $\alpha^{ij}(\omega)=\bar\alpha^{ij}(\omega)$. 
The pole at $\omega=0$ is associated with a delta function via
the Kramers-Kr\"onig relations, and is due to conservation of momentum (since any breaking of translations is 
assumed to be spontaneous). 
In the case that there is no scalar source associated with breaking of time reversal invariance, then we also have that $\sigma^{ij}(\omega)$, $\alpha^{ij}(\omega)=\bar\alpha^{ij}(\omega)$ and $\bar\kappa^{ij}(\omega)$ are all symmetric matrices.

We now define the incoherent current operator
\begin{align}
J_{inc}^i\equiv (\varepsilon+p)J^i+\rho T^i{}_t
=Ts J^i-\rho Q^i\,.
\end{align}
For the backgrounds we are considering we have $\bar J_{inc}^i= Ts\bar J^i$, which
is zero both when the $U(1)$ symmetry is not spontaneously broken and also for superfluids in the thermodynamically preferred phase. We also notice that
$\delta \bar J_{inc}^i=0$, showing that $\bar J_{inc}^i$ is an invariant quantity under the finite frequency boosts \eqref{boost}.
From \eqref{eefive} and \eqref{eefive2} we have
\begin{align}
G_{J^{i}_{inc}{(Q^j+\mu J^j)}}(\omega,{\bf 0})&=
G_{(Q^i+\mu J^i)J^{j}_{inc}}(\omega,{\bf 0})=0\,.
\end{align}
Furthermore, defining the incoherent conductivity via $\sigma^{ij}_{inc}(\omega)\equiv\frac{i}{\omega} G_{J^i_{inc}J^j_{inc}}(\omega)$ we have
\begin{align}
\sigma^{ij}_{inc}(\omega)=
(Ts)^2\sigma^{ij}(\omega) -2(Ts) \rho T\alpha^{ij}(\omega)+\rho^2 T\bar\kappa^{ij}(\omega)\,.
\end{align}

At this juncture we now assume that the $U(1)$ is unbroken (i.e. no superfluid). In this case since $\sigma^{ij}_{inc}(\omega)$
is a boost invariant quantity then we expect it to be a finite quantity at $\omega=0$. Continuing now with this
assumption it is convenient to define $[\sigma^{ij}_{inc}]_{DC}\equiv \sigma^{ij}_{{inc}}(\omega=0)$ and also
\begin{align}
\sigma_0^{ij}\equiv \frac{1}{(\varepsilon+p)^2 }[\sigma^{ij}_{inc}]_{DC}\,.
\end{align}

As we have already seen there are poles in the thermoelectric conductivity matrices, and hence associated delta functions which we suppress for the moment. If we now assume that the analytic structure of the Green's functions is such that we can write $\sigma(\omega)\to \frac{i}{\omega}x+y$, as $\omega\to 0$, where $x,y$ are constant matrices, then using
\eqref{poles} to get expressions for $\alpha$ and $\bar\kappa$ as $\omega\to 0$, as well
as demanding that the pole is absent in $\sigma_{{inc}}(\omega)$ we immediately deduce that we can write, as $\omega\to 0$,
\begin{align}\label{fstmrs}
\sigma^{ij}(\omega)&\to\left(\pi\delta(\omega)+ \frac{i}{\omega}\right)\frac{\rho^2}{\varepsilon+p}\delta^{ij}+\sigma_0^{ij}\,,\nn
T\bar\alpha^{ij}=T\alpha^{ij}(\omega)&\to\left(\pi\delta(\omega)+ \frac{i}{\omega}\right)\frac{\rho Ts}{\varepsilon+p}\delta^{ij}-\mu\sigma_0^{ij}\,,\nn
T\bar\kappa^{ij}(\omega)&\to\left(\pi\delta(\omega)+ \frac{i}{\omega}\right)\frac{(Ts)^2}{\varepsilon+p}\delta^{ij}+\mu^2\sigma_0^{ij},
\end{align}
and here we have included the delta functions.
This is our first main result.

Some simple corollaries now follow. We first recall that the electrical conductivity at zero total heat current can be expressed as
$\sigma_{\bar Q=0}(\omega)\equiv\sigma(\omega)-T{\alpha}(\omega)\bar\kappa(\omega)^{-1}{\bar\alpha}(\omega)$, while
the thermal conductivity at zero total electric current is given by $\kappa(\omega)\equiv\bar\kappa(\omega)-T{\bar\alpha}(\omega)\sigma(\omega)^{-1}{\alpha}(\omega)$. From \eqref{fstmrs} we deduce that $\sigma_{\bar Q=0}(\omega)$ and also $\kappa(\omega)$, if $\rho\ne 0$,
are both finite as $\omega\to 0$ with
\begin{align}\label{qzjz}
\sigma^{ij}_{\bar Q=0}(\omega)
\to  \frac{(\varepsilon+p)^2}{(Ts)^2}\sigma_0^{ij}\,,
\qquad
\kappa^{ij}(\omega)
\to \frac{(\varepsilon+p)^2}{\rho^2}\sigma_0^{ij}\,.
\end{align}
Furthermore, since $\alpha^{ij}(\omega)=\bar\alpha^{ij}(\omega)$, we can write
\begin{align}\label{qzeroexpw}
\sigma^{ij}_{inc}(\omega)=
(Ts)^2\sigma^{ij}_{\bar Q=0}(\omega) +[T\alpha\bar\kappa^{-1}\alpha(Ts-\rho\alpha^{-1}\bar\kappa)^2]^{ij}(\omega)\,,
\end{align}
and we note that the second term vanishes as $\omega\to 0$.

\section{Holography}
Within holography, phases with spontaneously broken translations are described by black holes
with planar horizons with a metric, gauge field and scalar which, generically, all depend periodically on all of the spatial directions. 
Such horizons also arise for ``holographic lattices", i.e. black hole solutions which are dual to field theories that have been deformed by operators which explicitly break spatial translations.

In both cases, following \cite{Donos:2015gia}, we briefly summarise how one can obtain the thermoelectric conductivity of the black hole horizon. 
To simplify the discussion we only consider background configurations that have vanishing magnetisation currents and moreover time-reversal invariance is not broken, either explicitly or spontaneously\footnote{More general discussions, including a careful treatment of transport currents, can be found in \cite{Donos:2015bxe,Donos:2017mhp}.}.
We will also assume we are not in a superfluid phase.

One first applies a suitable DC perturbation to the full black hole solution that is
linear in the time coordinate
and parametrised by DC sources $E_i$ and $\zeta_i$, which are taken to be constant throughout the bulk spacetime. It can then be shown that on the black hole horizon a subset of the perturbation must satisfy a Stokes flow for an auxiliary fluid, with sources $E_i$ and $\zeta_i$. Solving these Stokes equations gives 
local currents on the horizon, $Q^i_H$ and $J^i_H$, which depend periodically on the spatial coordinates.
Determining the zero modes of these currents, denoted by $\bar Q^i_H$ and $\bar J^i_H$, and relating them to $E_i$ and $\zeta_i$ we then obtain, by definition, the horizon DC conductivities $\sigma^{ij}_H$, $\alpha_H^{ij}$, $\bar\alpha_H^{ij}$ and
$\bar\kappa^{ij}_H$. In the absence of Killing vectors on the black hole horizon geometry, these will be uniquely defined and finite quantities. Since we are assuming that the background
is time-reversal invariant, $\sigma^{ij}_H$ and $\bar \kappa^{ij}_H$ are symmetric matrices and also
$\alpha^{ij}_H=\bar\alpha^{ji}_H$. 

In the case of holographic lattices, i.e. when the translations have been explicitly broken, all DC conductivities of the dual field theory will be finite and $\sigma^{ij}_H$, $\alpha_H^{ij}$, $\bar\alpha_H^{ij}$, $\bar\kappa^{ij}_H$ are equal to the associated DC conductivities 
$\sigma^{ij}_{DC}$, $\alpha_{DC}^{ij}$, $\bar\alpha_{DC}^{ij}$, $\bar\kappa^{ij}_{DC}$ of the dual field theory \cite{Donos:2015gia}.
This result follows after showing that the zero modes of the currents on the horizon, $\bar Q^i_H$ and $\bar J^i_H$, which are finite, are equal to the zero modes of the currents at the holographic boundary. 

Turning to the case that translations have been broken spontaneously, the DC conductivities 
of the dual field theory contain infinities due to the presence of Goldstone modes. Thus, $\sigma^{ij}_H$, $\alpha_H^{ij}$, $\bar\alpha_H^{ij}$, 
$\bar\kappa^{ij}_H$, which are finite, are certainly not equal to the $\sigma^{ij}_{DC}$, $\alpha_{DC}^{ij}$, $\bar\alpha_{DC}^{ij}$, $\bar\kappa^{ij}_{DC}$, the DC conductivities of the dual field theory. 
However, since the zero modes of the currents on the horizon $\bar Q^i_H$ and $\bar J^i_H$,
are finite and, moreover, they are still equal to the zero modes of the currents at the holographic boundary, this seems paradoxical.
The simple resolution is that the full linearised perturbation about the black hole solution, with sources parametrised by $E_i$ and $\zeta_i$ and regular at the black hole horizon, is no longer unique in the bulk spacetime. Indeed, when translations are broken spontaneously, by carrying out a coordinate transformation of the bulk solution, we can generate additional time dependent solutions that are regular at the horizon and without additional sources at the $AdS$ boundary, as we explain in more detail in
appendix \ref{nonu}.

Nevertheless, in the case that translations are broken spontaneously we know that there is a finite DC conductivity, namely
$[\sigma_{inc}]^{ij}_{DC}\equiv
\sigma^{ij}_{inc}(\omega\to 0)$, and this quantity can be obtained from a Stokes flow on the horizon. One applies a DC perturbation in which we source the incoherent current,
$J_{inc}^i$, but not the current $Q^i+\mu J^i$, and this is achieved\footnote{This can by seen by writing $\tilde J_A=M_{AB} J_B$, where $\tilde J_A=(J^{inc},Q+\mu J)$, $J_A=(J^i,Q^i)$  and deducing that the corresponding transformed sources are $\tilde s= (M^T)^{-1} s$ in order that $J^Ts=\tilde J^T \tilde s$. 
Furthermore, if we set $\zeta_i=-\frac{\rho}{Ts} E_i$ we have $\tilde s=(E/(Ts),0)$.} by taking $\zeta_i=-\frac{\rho}{Ts} E_i$. Solving the Stokes flow on the horizon with this source, one obtains a local incoherent current on the horizon, whose zero mode is
also the zero mode of the incoherent current in the boundary theory, $\bar J_{inc}$. 
Since we have
$\bar J^i_{inc}=\left((Ts)^2\sigma^{ij}_H -Ts \rho[T\alpha^{ij}_H+T\bar\alpha^{ij}_H]+\rho^2 T\bar\kappa^{ij}_H\right)E_j/(Ts)$ we deduce that
when translations are broken spontaneously\footnote{In the case of translationally invariant backgrounds, 
the horizon has Killing vectors and there is not a unique solution to the Stokes equations on the horizon. Specifically, we can have $v^i$ proportional
to a Killing vector on the horizon, with $p,w$ constant, in the notation of \cite{Donos:2015gia,Banks:2015wha}. However, this ambiguity drops out of the incoherent current on the horizon, $J^i_{Hinc}\equiv (Ts)J^i_H-\rho Q^i_H$, which in this setting is constant. Furthermore, applying sources with $\zeta_i=-\frac{\rho}{Ts} E_i$ and writing  $J^i_{Hinc}=[\sigma_{inc}]^{ij}_{DC}E_j/(Ts)$ we can obtain an expression for $[\sigma_{inc}]^{ij}_{DC}$. 
For example, for the general class of models considered in \cite{Donos:2015gia,Banks:2015wha}, we get 
$[\sigma_{inc}]^{ij}_{DC}=(Ts)^2\sqrt{g_0}g^{ij}_0Z_0$.
This gives an alternative approach to obtaining $[\sigma_{inc}]^{ij}_{DC}$ than that discussed in \cite{Davison:2015taa}.}
$[\sigma_{inc}]^{ij}_{DC}$ is given by
\begin{align}\label{dchorinc1}
[\sigma_{inc}]^{ij}_{DC}=
(Ts)^2\sigma^{ij}_H -Ts \rho[T\alpha^{ij}_H+T\bar\alpha^{ij}_H]+\rho^2 T\bar\kappa^{ij}_H\,.
\end{align}
In particular, we deduce that the 
DC conductivity for the incoherent current of the field theory
can be expressed in terms of the horizon DC conductivities, obtained from the solution to the Stokes flow on the horizon, combined with specific thermodynamic quantities
of the equilibrium black hole solutions, which also can be obtained from the horizon. This is the second main result of this paper.

We repeat that, in general, the individual horizon conductivities on the right hand side of 
\eqref{dchorinc1} are not the same as those of the boundary field theory. In particular, 
despite that fact that from \eqref{qzeroexpw} we have $[\sigma_{inc}]^{ij}_{DC}=(Ts)^2\sigma^{ij}_{\bar Q=0}(\omega\to 0)$
we do not have, in general, $\sigma^{ij}_{\bar Q=0}(\omega\to 0)=\sigma^{ij}_{\bar Q_H=0}$,
where $\sigma_{\bar Q_H=0} \equiv \sigma_H-T{\alpha}_H\bar\kappa_H^{-1}{\bar\alpha}_H$
is the horizon DC conductivity for vanishing zero mode of the horizon heat current.

To further clarify this point, it is illuminating to now consider the black hole horizon to be a small perturbation about a flat planar space, parametrised by a small number $\lambda$. It was shown in \cite{Donos:2015gia,Banks:2015wha} that the horizon conductivities 
$\sigma^{ij}_H$, $\alpha_H^{ij}$, $\bar\alpha_H^{ij}$, 
$\bar\kappa^{ij}_H$ are of order $\lambda^{-2}$ but 
$\sigma_{\bar Q_H=0}$ is of order 
$\lambda^0$, with order $\lambda$ corrections. As we explain in appendix \ref{pertlat},
by extending the results of \cite{Donos:2015gia,Banks:2015wha}
we can actually deduce that
\begin{align}\label{dcplat}
[\sigma_{inc}]^{ij}_{DC}=
{(Ts)^2}\sigma_{\bar Q_H=0}^{ij}({\lambda}) +\mathcal{O}(\lambda^2)\,,
\end{align}
and $[\sigma_{inc}]^{ij}_{DC}\ne {(Ts)^2}\sigma_{\bar Q_H=0}^{ij}({\lambda})$, in general.
If we let $T_c$ be the temperature for the phase transition that spontaneously breaks translations, then for temperatures just below $T_c$
the horizon will be a small deformation away from flat space, parametrised\footnote{Here we are assuming that the phase transition has mean field exponents, with the expectation value of the order parameter proportional to $(1-T/T_c)^{1/2}$. This implies that the horizon can be expanded in the same parameter about flat space.} by
$\lambda\sim (1-T/T_c)^{1/2}$. Since, by direct calculation as in footnote 4, the value of $[\sigma_{inc}]_{DC}$ for the translation invariant background for temperatures above $T_c$ is the same as
${(Ts)^2}\sigma_{\bar Q_H=0}({\lambda\to 0})$, 
we see that $[\sigma_{inc}]_{DC}$ is continuous as the temperature is lowered.

We can also consider $\lambda$ to parametrise a small explicit breaking of translations added to a system that
spontaneously breaks translations. In this case, all of the individual thermoelectric conductivity matrices of the dual field theory
are finite and equal to the horizon quantities. In this case it will only be near $T=T_c$ in which 
the horizon is a small deformation about flat space and then one can expand in either $\lambda$ or $(1-T/T_c)^{1/2}$.

\section*{Acknowledgements}
We thank Alexander Krikun for helpful discussions.
AD is supported by STFC grant ST/P000371/1.
JPG and TG are supported by the European Research Council under the European Union's Seventh Framework Programme (FP7/2007-2013), ERC Grant agreement ADG 339140. JPG is also supported by STFC grant ST/P000762/1, EPSRC grant EP/K034456/1, as a KIAS Scholar and as a Visiting Fellow at the Perimeter Institute.

\appendix
\section{Non-thermodynamically preferred phases}\label{nonthermpref}
If we consider the system in thermal equilibrium, but do not assume that we have minimised the action with respect
to the size and shape of the spontaneously formed lattice, as in \cite{Amoretti:2017frz},
then the formulas in the text are modified slightly. 
It is helpful to introduce the symmetric matrix 
$m^{ij}$ defined by
\begin{align}
m^{ij}\equiv \left(\varepsilon-\mu\rho\right)\delta^{ij}+t^{ij}\,,
\end{align}
so that for the thermodynamically preferred branches we have $m^{ij}=Ts\delta^{ij}$.

Equations \eqref{eefive},\eqref{eefive2} get modified to
 \begin{align}\label{eefiveap}
G_{J^{i}{(Q^j+\mu J^j)}}(\omega,{\bf 0})&=G_{(Q^i+\mu J^i){J}^{j}}(\omega,{\bf 0})={\rho}\,\delta^{ij}\,,\nn
G_{{(Q^i+\mu J^i)}{(Q^j+\mu J^j)}}(\omega,{\bf 0})&=[m+\mu\rho]^{ij}\,\,.
\end{align}
This implies that \eqref{poles} should be changed to
\begin{align}\label{polesap}
\mu\,\sigma^{ij}(\omega)+T{\alpha}^{ij}(\omega)=&\frac{i{\rho}}{\omega}\,\delta^{ij}\,,\nn
\mu T\bar\alpha^{ij}(\omega)+T\bar{\kappa}^{ij}(\omega)=&\frac{i}{\omega}m^{ij}\,,
\end{align}
and ${\alpha}^{ij}(\omega)=\bar\alpha^{ij}(\omega)$.
The definition of the incoherent current operator is modified to
\begin{align}
J_{inc}^i\equiv [mJ]^i-\rho Q^i\,.
\end{align}
From \eqref{eefiveap} we have
\begin{align}
G_{J^{i}_{inc}{(Q^j+\mu J^j)}}(\omega,{\bf 0})&=
G_{(Q^i+\mu J^i)J^{j}_{inc}}(\omega,{\bf 0})=0\,,
\end{align}
and the incoherent conductivity, $\sigma^{ij}_{inc}(\omega)\equiv\frac{i}{\omega} G_{J^i_{inc}J^j_{inc}}(\omega)$, is given by
\begin{align}
\sigma^{ij}_{inc}(\omega)=[m\sigma(\omega)m]^{ij}-\rho[T\alpha(\omega) m+m T\alpha(\omega)]^{ij}+\rho^2T\bar\kappa^{ij}(\omega)\,.
\end{align}

Writing $\sigma(\omega)\to \frac{i}{\omega}x+y$, as $\omega\to 0$, where $x,y$ are constant matrices, as in the text, we deduce that
\begin{align}
\sigma^{ij}(\omega)&\to \left(\pi\delta(\omega)+ \frac{i}{\omega}\right){\rho^2}[(m+\mu\rho)^{-1}]^{ij}+\sigma_0^{ij}\,,\nn
T\bar\alpha^{ij}=T\alpha^{ij}(\omega)&\to\left(\pi\delta(\omega)+ \frac{i}{\omega}\right){\rho}[m(m+\mu\rho)^{-1}]^{ij}-\mu\sigma_0^{ij}\,,\nn
T\bar\kappa^{ij}(\omega)&\to \left(\pi\delta(\omega)+ \frac{i}{\omega}\right)[m^2(m+\mu\rho)^{-1}]^{ij}+\mu^2\sigma_0^{ij}\,,
\end{align}
where
\begin{align}
\sigma_0^{ij}\equiv [(m+\mu\rho)^{-1}[\sigma_{inc}]_{DC}(m+\mu\rho)^{-1}]^{ij}\,,
\end{align}
and $[\sigma_{inc}]^{ij}_{DC}=\sigma^{ij}_{{inc}}(\omega=0)$.

In the holographic setting, in order to get $[\sigma_{inc}]^{ij}_{DC}$ we can solve the Stokes flow on the horizon with
the following constraint on the sources: $\zeta_i=-\rho (m^{-1}E)_i$. This leads to
\begin{align}\label{dchorinc2ap}
[\sigma_{inc}]^{ij}_{DC}&=
[m\sigma_Hm]^{ij}-\rho T[\bar \alpha_H m+m \alpha_H]^{ij}+\rho^2T\bar\kappa^{ij}_H\,.
\end{align}
Once again we can obtain $[\sigma_{inc}]^{ij}_{DC}$ from horizon data supplemented with thermodynamic properties of the background.
It is worth noting that here, in contrast to \eqref{dchorinc1}, not all of the thermodynamic quantities can be obtained directly from the horizon.

The above formulae simplify somewhat for the special case of spatially isotropic phases in which  
all of the horizon conductivities are proportional to the identity matrix and furthermore $t^{ij}=t\delta^{ij}$, 
so that $m^{ij}=(Ts+w+t)\delta^{ij}$. In this setting
we have
\begin{align}
\sigma(\omega)&\to \left(\pi\delta(\omega)+ \frac{i}{\omega}\right)\frac{\rho^2}{\varepsilon+t}+\sigma_0\,,\nn
T\bar\alpha=T\alpha(\omega)&\to\left(\pi\delta(\omega)+ \frac{i}{\omega}\right)\frac{\rho(Ts+w+t)}{\varepsilon+t}-\mu\sigma_0\,,\nn
T\bar\kappa(\omega)&\to \left(\pi\delta(\omega)+ \frac{i}{\omega}\right)\frac{(Ts+w+t)^2}{\varepsilon+t}+\mu^2\sigma_0\,,
\end{align}
where
\begin{align}\label{simplesignought}
\sigma_0=\frac{1}{(\varepsilon+t)^2}[\sigma_{inc}]_{DC}\,.
\end{align}
In the holographic setting, for this special case, we can write
\begin{align}\label{dchorinc2ap2}
[\sigma_{inc}]_{DC}&=(Ts+w+t)^2\sigma_{\bar Q_H=0}+T\alpha^2\bar\kappa^{-1}(Ts+w+t-\rho\alpha^{-1}\bar\kappa)^2\,.
\end{align}
For the special case of an isotropic Q-lattice with $d$ spatial dimensions we can be more explicit and this will allow us to recover
the result of \cite{Amoretti:2017frz} for $[\sigma_{inc}]_{DC}$ who used a different approach. Using the same
notation as in section 4.1 of \cite{Banks:2015wha} the breaking of translations is specified by a matrix $\mathcal{D}_{ij}$, which for an isotropic lattice can be written as $\mathcal{D}_{ij}\equiv \mathcal{D}\delta_{ij}$. Substituting
the results of \cite{Banks:2015wha} into \eqref{dchorinc2ap2} then easily gives
\begin{align}\label{dchorinc2ap3}
[\sigma_{inc}]_{DC}&=(Ts+w+t)^2(\frac{s}{4\pi})^{(d-2)/d}Z_H+\frac{4\pi \rho^2(w+t)^2}{s\mathcal{D}}\,.
\end{align}
Combining this with \eqref{simplesignought} and setting $d=2$, we obtain equation (74) of \cite{Amoretti:2017frz} 
after identifying $w+t$ with $-2K$ in their notation.

\section{Bulk non-uniqueness}\label{nonu}

We consider a holographic theory describing a relativistic quantum field theory at finite temperature defined on flat spacetime. The system is held at constant chemical potential, $\mu$, with respect to an abelian global symmetry and we will also allow for the possibility for additional deformations of the Hamiltonian by an uncharged scalar operator $\mathcal{O}_\phi$ that
is parameterised by the constant $\phi_{s}$.

We consider the following bulk coordinate transformations
\begin{align}\label{coordtrans}
x^i\to x^i-u^i (t+S(r))\,,\qquad t\to t- v_i x^i\,,
\end{align}
as well as a gauge transformation with parameter $\Lambda=\mu w_i x^i$, where $u^i$, $v_i$ and $w_i$ are all constant vectors. Here $S(r)$ is a function of the holographic radial coordinate such that $S(r)=\frac{\ln r}{4\pi T}+...$ near the horizon, located at $r\to 0$, 
and $S(r)\rightarrow 0$ as one approaches the AdS boundary located at $r\rightarrow\infty$. 
This transformation adds the following boundary sources:
\begin{align}
\delta g_{ti}=v_i-\delta_{ij}u^j,\qquad \delta A_i=\mu (w_i-v_i)\,.
\end{align}
In particular, setting $v_i=\delta_{ij}u^j$ and $w_i=v_i$ gives a source free transformation that is regular at the black hole horizon. This means that, demanding a given set of sources on the AdS boundary combined with regularity at the black hole horizon, does not lead to a unique solution to the bulk equations of motion.

If we take the parameters to be infinitesimal perturbations we also deduce the following transformations on the currents in the boundary field theory:
\begin{align}
\delta \langle J^{i}\rangle =&-tu^k\partial_k \langle J^{i}\rangle+u^i  \langle J^{t}\rangle\,,\nn
\delta \langle T^i{}_t\rangle =&-tu^k\partial_k \langle T^i{}_t\rangle +u^i\langle T^t{}_t\rangle - u^j\langle T^i{}_j\rangle\,.
\end{align}
If we consider the zero modes we have
\begin{align}\label{xmzms}
\delta \langle \bar J^{i}\rangle =&u^i  \rho\,,\nn
\delta \langle \bar T^i{}_t\rangle =&-u^j(\varepsilon\delta_j^i+t^{ik}\delta_{jk})\,,
\end{align}
and also $\delta \langle \bar Q^i \rangle=-\delta \langle \bar T^i{}_t\rangle-\mu \delta \langle \bar J^{i}\rangle=u^j m^{ik}\delta_{jk}$.\footnote{Note that on the thermodynamically preferred branch we have $\delta \langle \bar J^{i}\rangle =u^i  \rho$, 
$\delta \langle \bar T^i{}_t\rangle =-u^i(\varepsilon+p)$ and $\delta \bar Q^i=u^i Ts $} We see that both $\langle \bar J^{i}\rangle$ and $\langle \bar Q^i \rangle$ are changed by this transformation (when $\rho\neq0$). In particular, this means that the DC thermoelectric conductivity matrix is not well defined (when $\rho\neq0$). Note, however, that $\delta \langle \bar{J}_{inc}^i \rangle=0$. 

This non-uniqueness of bulk solutions (and of the currents $\langle \bar J^{i}\rangle$ and $\langle \bar Q^i \rangle$) means that we must be careful when calculating the DC conductivities. A solution to the perturbed equations parametrised by $E_i,\zeta_i$
may be found for which the DC current response is finite everywhere (including at the horizon and at the boundary), but when this is not the unique solution for the current, the associated DC conductivity will not be well-defined. The conclusion is that when calculating DC conductivities, it is important to first establish that the associated current response is uniquely defined.

\section{Perturbative Lattice}\label{pertlat}

We follow the analysis and notation of \cite{Donos:2015gia,Banks:2015wha} which focussed on Einstein-Maxwell-dilaton theory
with Lagrangian density $\mathcal{L}=R-V(\phi)-\tfrac{1}{4}Z(\phi) F^2-\tfrac{1}{2}(\partial\phi)^2$. 
For the black holes of interest, which preserve time reversal invariance, 
we assume that at the black hole horizon we can expand about a flat geometry using a perturbative parameter $\lambda$:
\begin{align}\label{eq:pert_exp}
g_{(0)}{}_{ij}&=g\,\delta_{ij}+\lambda\,h^{(1)}_{ij}+\cdots\,, \qquad
{Z^{(0)}a_{t}^{(0)}}=a+\lambda\, a_{(1)}+\cdots\,,\nn
\phi^{(0)}&=\psi_{(0)}+\lambda\,\psi_{(1)}+\cdots\,,
\qquad
Z^{(0)}=z_{(0)}+\lambda\,z_{(1)}+\cdots\,,
\end{align}
with $a$, $z_{(0)}$, $\psi_{(0)}$ and $g$ being constant and the sub-leading terms are functions of, generically, all of the spatial coordinates $x^i$ and they respect the lattice symmetry. 
We can calculate the entropy density $s=\oint s_H$ and the charge density $\rho=\oint \rho_H$ on the horizon using
\begin{align}\label{seetwo}
s_H&\equiv 4\pi \sqrt{g_{(0)}}=4\pi g^{d/2}(1+\lambda\frac{h^{(1)}}{2g}+\cdots),\nn
 \rho_H&\equiv \sqrt{g_{(0)}}Z^{(0)}a_t^{(0)}=a g^{d/2}(1+\lambda(\frac{h^{(1)}}{2g}+\frac{a_{(1)}}{a})+\cdots)\,,
\end{align}
where $h^{(1)}=\delta^{ij}h^{(1)}_{ij}$ and $d$ is the number of spatial dimensions.
We thus\footnote{Note that if preferred, one could absorb the zero modes of all the sub-leading terms in \eqref{eq:pert_exp} into the leading terms, $g$, $a$, etc. and then $s$ and $\rho$ could be expressed in terms of the resummed, constant, leading terms plus
corrections that would be of order $\mathcal{O}(\lambda^2)$ (since, as we see from \eqref{seetwo}, the $\mathcal{O}(\lambda)$ pieces would vanish when integrated over the
spatial coordinates).} have $s=4\pi g^{d/2}+\mathcal{O}(\lambda)$ and
$\rho=a g^{d/2}+\mathcal{O}(\lambda)$.

As shown in \cite{Donos:2015gia,Banks:2015wha}, we can solve the horizon constraint equations perturbatively in $\lambda$ using the following
expansion:
\begin{align}
v^i&=\frac{1}{\lambda^{2}}\,v_{(0)}^i+\frac{1}{\lambda}\,v_{(1)}^i+v_{(2)}^i+\cdots\,,\qquad
w=\frac{1}{\lambda}\,w_{(1)}+w_{(2)}+\cdots\,,\notag\\
p&=\frac{1}{\lambda}\,p_{(1)}+p_{(2)}+\cdots\,,
\end{align}
where $v_{(0)}^i$ is constant. In \cite{Donos:2015gia,Banks:2015wha}, this then yields a solution for the horizon DC thermoelectric conductivities $\sigma_H^{ij}$, $\alpha_H^{ij}$, $\bar{\alpha}_H^{ij} $ and $\bar{\kappa}_H^{ij}$, which all have leading order behaviour of order $1/\lambda^2$.

We can now make some additional observations. We can calculate the zero modes of the electric and heat
current as follows.
\begin{align}\label{jcur}
\bar{J}^i_{(0)}&\equiv \oint\sqrt{g_{(0)}}Z^{(0)}(a_t^{(0)}v^i+g^{ij}_{(0)}(\partial_jw+E_j))
=\oint \rho _Hv^i+\mathcal{O}(\lambda^0)\,,\nn
&=\oint \rho_H(\frac{1}{\lambda^{2}}\,v_{(0)}^i+\frac{1}{\lambda}\,v_{(1)}^i)+\mathcal{O}(\lambda^0)
=(\frac{1}{\lambda^{2}}\,\rho v_{(0)}^i+\frac{1}{\lambda}\,\rho \bar{v}_{(1)}^i)+\mathcal{O}(\lambda^0)\,,\nn
&=\rho \bar{v}^i+\mathcal{O}(\lambda^0)\,.
\end{align}
Similarly,
\begin{align}\label{qcur}
\bar{Q}^i_{(0)}&\equiv 4\pi T\oint\sqrt{g_{(0)}}v^i\,,\nn
&=T s\bar{v}^i+\mathcal{O}(\lambda^0)\,.
\end{align}
We thus have $\rho \bar{Q}^i_{(0)}=sT \bar{J}^i_{(0)}+\mathcal{O}(\lambda^0)$, which means that $\rho T (\bar{\kappa}_H^{ij}\zeta_j+\bar{\alpha}_H^{ij}E_j)=sT (\sigma_H^{ij}E_j+T\alpha_H^{ij}\zeta_j)+\mathcal{O}(\lambda^0)$. Since this holds for arbitrary $E_j$ and $\zeta_j$, we must have:
\begin{align}\label{horrel}
\sigma_H^{ij}&=\frac{\rho}{s} \bar{\alpha}_H^{ij}+\mathcal{O}(\lambda^0)\,,\nn
\alpha_H^{ij}&=\frac{\rho}{sT} \bar{\kappa}_H^{ij}+\mathcal{O}(\lambda^0)\,.
\end{align}
The Onsager relations for this time-reversal invariant background imply that $\alpha_H^{ij}=\bar{\alpha}_H^{ji}$, $\sigma_H^{ij}=\sigma_H^{ji}$ and $\bar{\kappa}_H^{ij}=\bar{\kappa}_H^{ji}$, and so:
\begin{align}\label{seesev}
\sigma_H^{ij}&=\frac{\rho^2}{s^2T} \bar{\kappa}_H^{ij}+\mathcal{O}(\lambda^0)\,,\nn
\alpha_H^{ij}&= \bar{\alpha}_H^{ij}+\mathcal{O}(\lambda^0)=\frac{\rho}{sT} \bar{\kappa}_H^{ij}+\mathcal{O}(\lambda^0)\,.
\end{align}
We can use this to confirm that
\begin{align}
\sigma_{\bar Q_H=0}&\equiv \sigma_H-T\alpha_H\bar{\kappa}_H^{-1}\bar{\alpha}_H=\mathcal{O}(\lambda^0)\,,\nn
\kappa_{H}&\equiv \bar{\kappa}_H-T\bar{\alpha}_H\sigma_H^{-1}\alpha_H=\mathcal{O}(\lambda^0)\,.
\end{align}
Furthermore, using \eqref{seesev} we find
\begin{align}
\rho\bar \alpha_H^{-1}\bar{\kappa}_H=sT+\mathcal{O}(\lambda^2)\,,\qquad
\rho\bar\kappa_H\alpha_H^{-1}=sT+\mathcal{O}(\lambda^2)\,.
\end{align}

Now, from \eqref{dchorinc1} we can write the DC conductivity for the incoherent current as
\begin{align}\label{dchorinc12}
[\sigma_{inc}]_{DC}=
&(Ts)^2\sigma_{\bar Q_H=0} +
\frac{1}{2}T\alpha_H\bar\kappa_H^{-1}\bar \alpha_H(Ts-\rho\bar \alpha_H^{-1}\bar\kappa_H)^2
+
\frac{1}{2}T(Ts-\rho\bar\kappa_H\alpha_H^{-1})^2\alpha_H\bar\kappa_H^{-1}\bar \alpha_H\nn
&+\frac{1}{2}T\rho^2(\bar\alpha_H-\alpha_H)\bar\alpha_H^{-1}\bar\kappa_H-\frac{1}{2}T\rho^2\bar\kappa_H\alpha^{-1}_H(\bar \alpha_H-\alpha_H)\,.
\end{align}
The above results then allow us to conclude that
\begin{align}
[\sigma_{inc}]_{DC}=
{(Ts)^2}\sigma_{\bar Q_H=0}(\lambda)+\mathcal{O}(\lambda^2)\,.
\end{align} 
Note that we don't expect that \eqref{horrel} will continue to hold for higher orders in $\lambda$ (it arises from the very special form of the last line in \eqref{jcur} and \eqref{qcur}). 
Thus, in general, $[\sigma_{inc}]_{DC}\neq{(Ts)^2}\sigma_{\bar Q_H=0}$ .


\providecommand{\href}[2]{#2}\begingroup\raggedright\endgroup

\end{document}